\documentclass[%
 reprint, %draft,
 amsmath,amssymb,
 aps,
]{revtex4-2}

\usepackage{graphicx}% Include figure files
\usepackage{dcolumn}% Align table columns on decimal point
\usepackage{bm}% bold math
\usepackage{hyperref}% add hypertext capabilities
\usepackage{stmaryrd}

\newcommand{\fs}[1]{\left< #1\right>}
\newcommand{\favre}[1]{\langle\!\langle  #1\rangle\!\rangle}

\begin{document}

\preprint{APS/123-QED}

\title{First-Principles Explanation of the Drift Configuration Dependence of the Radial Electric Field and High-Confinement Access in Tokamaks}

\author{B. J. Frei$^{1}$}
\author{R. Bilato$^{1}$}
\author{O. Grover$^{1}$}
\author{W. Zholobenko$^{1}$}
\author{C. Angioni$^{1}$}
\author{M. Bergmann$^{1}$}
\author{P. Ulbl$^{1}$}
\author{F. Jenko$^{1}$}
\author{the ASDEX Upgrade Team$^{2}$}

\affiliation{
 $^{1}$  Max-Planck Institute for Plasma Physics, Boltzmannstr. 2, Garching, D-85748, Germany
}
\affiliation{$^{2}$ See author list of H. Zohm et al 2024 Nucl. Fusion https://doi.org/10.1088/1741-4326/ad249d}

\date{\today}

\begin{abstract}
The origin of the difference in the high-confinement (H-mode) power threshold between favorable and unfavorable drift configurations in tokamaks—experimentally linked to a deeper radial electric field ($E_r$) well in the former—remains unresolved. Using first-principles gyrokinetic simulations of edge and scrape-off-layer turbulence in ASDEX Upgrade, we show that turbulence-driven poloidal flows generate this deeper $E_r$ well in the favorable configuration through enhanced nonlinear turbulence–mean flow energy transfer. This transfer is significantly weaker in the unfavorable case, yielding a shallower $E_r$ well, while turbulence intensity is simultaneously higher. Within the turbulence–flow shear suppression paradigm, the combination of stronger shear and reduced turbulence facilitates H-mode access in the favorable configuration. These results provide the first validated, self-consistent full-$f$ gyrokinetic explanation of how drift configuration controls the nonlinear dynamics of profiles, $E_r$, flows, and turbulence, thereby setting the H-mode power threshold.
\end{abstract}

%\keywords{Suggested keywords}Use showkeys class option if keyword
                              %display desired
\maketitle

In tokamaks, the direction of the ion $\nabla B$ drift (relative to the active X-point) defines favorable (\textit{fav}) and unfavorable (\textit{unfav}) configurations. This change of direction significantly influences the H-mode power threshold, with \textit{unfav} discharges generally requiring more than twice the threshold power than \textit{fav} ones~\cite{ryter1996,mckee2009,ryter2016,cziegler2017,plank2023}. This difference observed in multiple devices~\cite{carlstrom2002,schirmer2006,vermare2021,plank2023,yan2023} correlates with a deeper radial electric field ($E_r$) well near the separatrix in \textit{fav} before the transition from low-confinement (L-mode) to H-mode. A threshold in $E_r$ at the L-H transition has been identified in \textit{fav} at ASDEX Upgrade (AUG) \cite{plank2023b}, while it remains unclear in \textit{unfav}. Understanding the mechanism underlying this difference in $E_r$ is crucial for ITER and future fusion reactors, as it determines the heating power required and limits operational flexibility.
 
Recent experiments at AUG comparing \textit{fav} and \textit{unfav} discharges at matched parameters \cite{plank2023} show that neoclassical (NC) effects alone cannot explain the observed $E_r$ differences, highlighting a significant role for non-NC physics. Proposed mechanisms include ion orbit losses~\cite{versloot2011,brzozowski2019,zhu2023}, scrape-off-layer flows~\cite{labombard2005}, and nonlinear turbulence–mean flow interactions driven by the Reynolds stress (RS) \cite{hidalgo1999,diamond2005,fedorczak2012,fedorczak2013,cziegler2017,schmid2017,grover2024}. In particular, turbulence–mean flow energy transfer is highly sensitive to poloidal asymmetries in the RS, which can be driven by X-point geometry, flow shear, and radial transport \cite{schmid2017,manz2018}. Reduced models incorporating such asymmetries capture qualitative experimental trends \cite{grover2024}, but a predictive, self-consistent, and nonlinear description of the interplay between equilibrium profiles, $E_r$, flows, and turbulence remains to be established. Existing studies—based on linear, simplified or quasi-linear approaches in simplified geometries—are not able to capture correctly the non-local coupling and nonlinear feedback between turbulence, mean flows, and $E_r$ in realistic diverted configurations. In addition, edge turbulence requires a GK treatment. Developing such a self-consistent and GK description is fundamental to determining whether these coupled effects fully account for the differences between \textit{fav} and \textit{unfav} and to quantifying their respective contributions.
 
In this Letter, we present the first gyrokinetic (GK) study of \textit{fav} and \textit{unfav} discharges at AUG \cite{plank2023} using validated first-principles simulations of edge-SOL turbulence. We show that stronger \textit{nonlinear} generation of poloidal flows near the separatrix explains the deeper $E_r$ well in \textit{fav}. An evolution equation for the \textit{flux-surface–averaged} (FSA) poloidal kinetic energy reveals that these poloidal flows are driven by radially localized asymmetries in the total velocity stress $\Pi$, a full-$f$ generalization of RS. These asymmetries arise from the orientation of $\mathbf{E}\times\mathbf{B}$ flow relative to the X-point on the low-field side (LFS), producing a radial gradient of $\Pi$ that enhances turbulence–flow energy transfer in \textit{fav}. Meanwhile, turbulence intensity is stronger in \textit{unfav}. Overall, the higher H-mode power threshold in \textit{unfav} might result from both weaker turbulence–mean flow energy transfer and stronger turbulence levels.

\textit{Experiments at AUG.} Dedicated deuterium experiments at AUG \cite{plank2023} compared a pair of lower single-null (LSN) discharges, \#36983 (\textit{fav}) and \#37375 (\textit{unfav}), to investigate the effect of the ion $\nabla B$ drift direction on $E_r$. Both discharges were performed under matched conditions: line-averaged electron density $n_e \simeq 2.7\times10^{19}$~m$^{-3}$ (lower density branch), $600$~kW of ECRH in L-mode. The on-axis toroidal magnetic field and plasma current were $\lvert B_\phi\rvert=2.5$~T and $\lvert I_p\rvert=0.8$~MA. Measurements of $E_r$ were obtained from Doppler back-scattering (DBS), while the electron temperature ($T_e$) and $n_e$ were reconstructed via integrated data analysis (IDA) \cite{fischer2010}. The main ion temperature ($T_i$) was assumed to be equal to that of the impurity species measured by charge-exchange recombination spectroscopy (CXRS). The deuterium toroidal ($U_{\phi i}$) and poloidal ($U_{\theta i}$) velocities were inferred from impurity velocity measurements using the NC code \texttt{NEOART} \cite{peeters2000}.

\textit{Turbulence Gyrokinetic Modelling.} We model these two discharges in the L-mode phase using the spectral version of the \texttt{GENE-X} code \cite{michels2021,frei2025vspec}, which solves the full-$f$, electromagnetic, long-wavelength, and collisional GK Vlasov equation with X-points. The details of the numerical setup is reported in \ref{sec:endmatter}. The magnetic equilibrium at $t = 2.7$~s is considered for the \textit{fav} discharge (\#36983) in the L-mode phase (prior to the L-H transition triggered at $2.78$~s with a power threshold of $1.1$~MW). To minimize differences between the two configurations in the simulations, the \textit{unfav} equilibrium (\#37375) is generated by reversing the direction of $B_\phi$ from \#36983, changing the direction of the ion $\nabla B$ drift and the magnetic helicity. A density source is introduced near the separatrix to mimic neutral ionization. To reach a quasi-steady state, both simulations were run until $t \sim 2.5$~ms. In quasi-steady state, the total power crossing the separatrix is approximatively $0.5$~MW in both cases.  A time window of $0.1$~ms is employed to perform the statistics.

\begin{figure}
\includegraphics[scale=0.41]{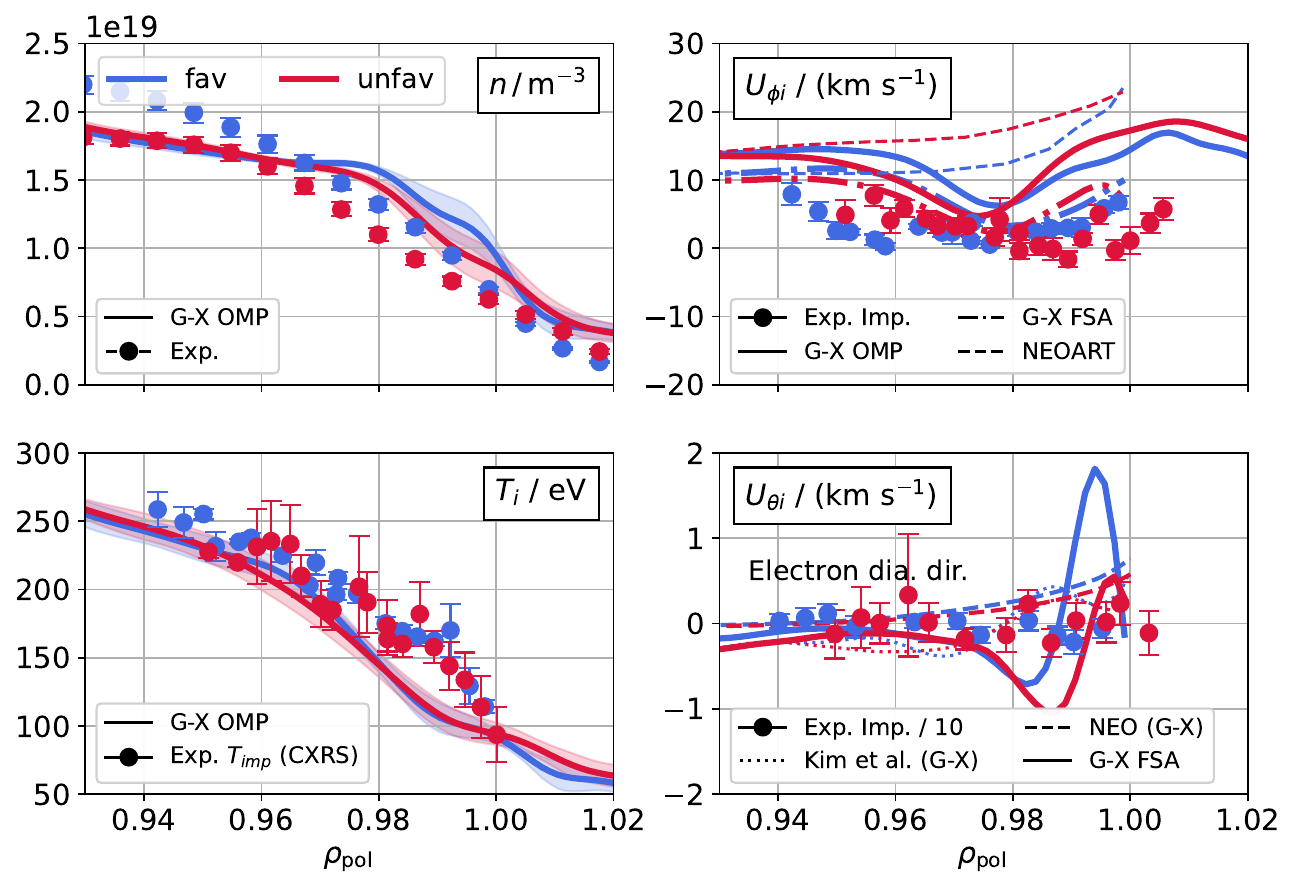} 
\caption{\label{fig:1} \textbf{Left:} OMP $n_e$ (top) and $T_i$ (bottom) from \texttt{GENE-X} (\texttt{G-X}, solid) in \textit{fav} (blue) and \textit{unfav} (red). Shaded areas denote standard deviations (time and toroidal averages). Experimental data \cite{plank2023} are shown as circles. \textbf{Right:} OMP and FSA $U_{\phi i}$ (top) and $U_{\theta i}$ (bottom). CXRS impurity ($U_{\phi i}$ and $U_{\theta i}$) and \texttt{NEOART} ($U_{\phi i}$) main ion velocities are included \cite{plank2023}. NC predictions for $U_{\theta i}$ from Kim \textit{et al.}~\cite{kim1991} and \texttt{NEO}~\cite{belli2008} are shown (dotted, dashed) using \texttt{GENE-X} profiles. The sign of \textit{unfav} $U_{\theta i}$ is reversed for comparison.}
\end{figure}

\textit{Validation of Simulation Results.} The left column of Fig.~\ref{fig:1} compares the outboard midplane (OMP) profiles (time and toroidally averaged) of $n_e$ and $T_i$ obtained from \texttt{GENE-X} with experimental measurements \cite{plank2023}. Overall, good agreement with experiments is observed in both configurations, albeit with slightly lower $T_i$ gradients near the separatrix. Most importantly, the simulations reproduce the experimental finding that the profiles are nearly insensitive to the ion $\nabla B$ drift direction \cite{plank2023}, although $n_e$ is slightly higher in \textit{fav} near $\rho_{\mathrm{pol}} \sim 0.99$. Similar observations (but not shown) can be made for $T_e$, with $T_i \sim 1.3 T_e$ at the separatrix.  

The right column of Fig.~\ref{fig:1} displays the OMP and FSA profiles of the deuterium toroidal (self-generated) and poloidal velocities. In \texttt{GENE-X}, the total (NC and turbulence-driven) ion velocity ($\bm U_{ i}$) is calculated from the zeroth-order velocity moment of the guiding-center drift corrected by a magnetization term \cite{dif2009,hazeltine2013}: $\bm U_i  =  \int dW \dot{\mathbf{R}} f_i / n_i + c \nabla \times \mathbf{m}_i  / (q_i n_i)$, with $\dot{\mathbf{R}}$ the guiding-center equation of motion, and $\mathbf{m}_i = - P_{\perp i} \mathbf{b} / B$ the classical magnetization. Then, $U_{\phi i}  = \bm U_{i} \cdot  \hat{\bm e}_\phi \simeq \bm U_{i} \cdot \bm b$ (with $\bm b= \bm B / B$) and $U_{\theta i} =  \bm U_{i} \cdot \bm \hat e_\theta$. As with $n_e$, $T_i$, $T_e$ (not shown)  $U_{\phi i}$ shows little dependence on the $\nabla B$ drift direction and is close to experiments \cite{plank2023}. Remarkably, the poloidal velocity $U_{i\theta}$ exhibits a clear difference: its maximal magnitude is much larger in \textit{fav} than in \textit{unfav} inside the $E_r$ well located near $\rho_{\mathrm{pol}} \sim 0.995$. In both cases, $U_{\theta i}$ is directed in the ion diamagnetic direction (in agreement with NC predictions) inside the edge ($\rho_{\mathrm{pol}} \lesssim 0.98$) and reverses toward the electron diamagnetic direction near the separatrix.

To assess the NC level of $U_{\theta i}$, we compare the \texttt{GENE-X} results with the NC prediction of Kim~\cite{kim1991} and \texttt{NEO} \cite{belli2008} calculations using the \texttt{GENE-X} profiles as inputs. Fig.~\ref{fig:1} demonstrates that $U_{\theta i}$ is consistent with NC close to the separatrix only in \textit{unfav}. On the other hand, the amplitude of $U_{\theta i}$ is significantly larger in the vicinity of the separatrix in \textit{fav}, indicating the presence of additional, non-NC mechanisms enhancing $U_{\theta i}$ in \textit{fav}.

\begin{figure}
\includegraphics[scale=0.43]{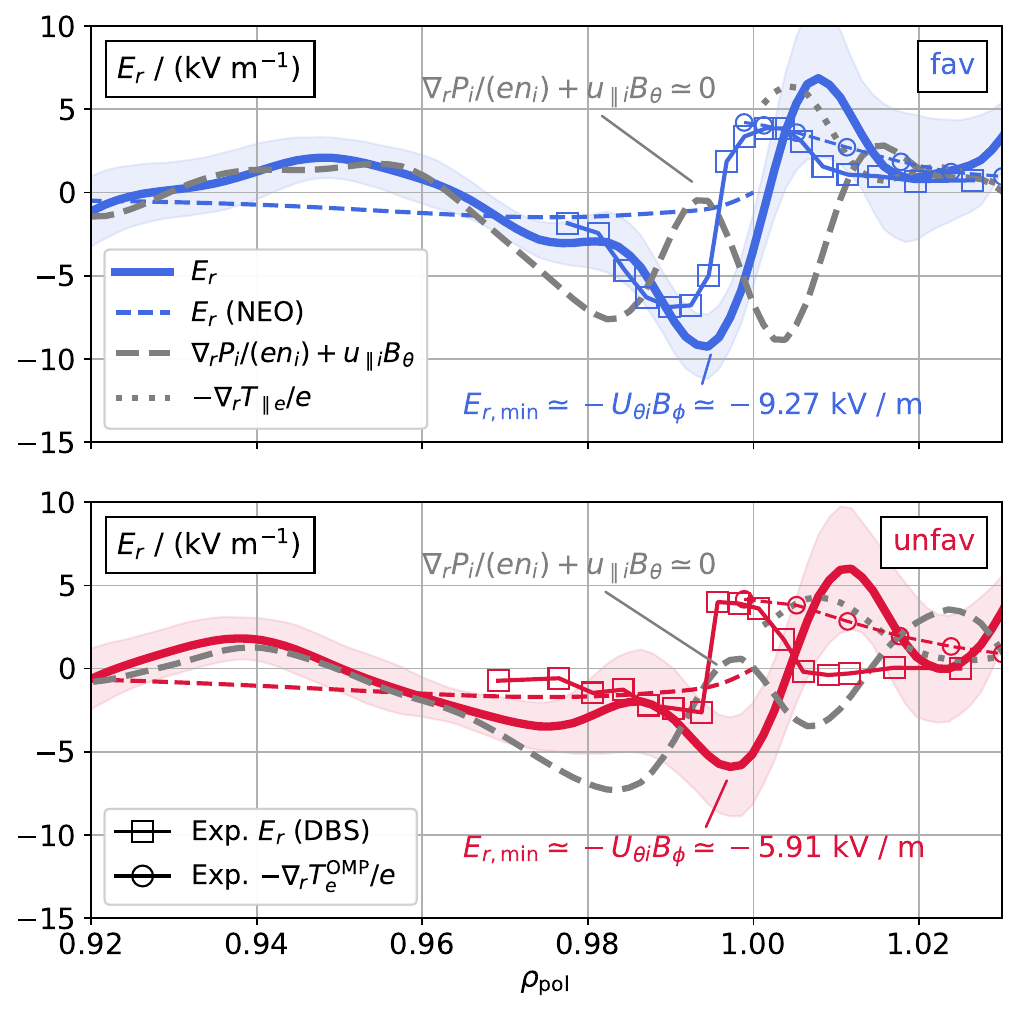}
 \caption{\label{fig:2} OMP $E_r$ from \texttt{GENE-X} (solid) in \textit{fav} (top) and \textit{unfav} (bottom). Shading indicates standard deviation. DBS measurements \cite{plank2023} are shown as squares with radial uncertainty $\Delta \rho_{\mathrm{pol}} \simeq 0.014$. NC $E_r$ from \texttt{NEO} (using \texttt{GENE-X} profiles) is shown (dashed). In both cases $E_{r,\min}\simeq -U_{i\theta}B_\phi$. SOL estimates from experiment ($-\nabla T_e^{\mathrm{OMP}}/e$, circles) and GK theory ($-\nabla T_{\parallel e}/e$ \cite{frei2025}, dotted) are also shown.}
\end{figure}

\textit{Radial Electric Field.} In \texttt{GENE-X}, $E_r$ is computed directly from the negative radial gradient of the electrostatic potential, $\phi_1$, obtained from the GK quasi-neutrality (QN) equation \cite{frei2025}. Fig.~\ref{fig:2} compares the simulated OMP $E_r$ profiles with experimental DBS measurements~\cite{plank2023} and NC \texttt{NEO} results (using \texttt{GENE-X} profiles). The simulations reproduce the overall experimental $E_r$ behavior in both the edge and SOL, with good agreement with experiments (within statistical and experimental uncertainties \cite{plank2023}). Enhanced geodesic acoustic mode (GAM) \cite{conway2021} activity is also found around the $E_r$ well in \textit{unfav}, being much smaller in \textit{fav} (see Fig.~\ref{fig:7}). Consistent with experiments, a pronounced difference arises near the separatrix: the $E_r$ well in \textit{fav} is significantly deeper than in \textit{unfav}. This deviation occurs despite nearly identical $n_e$, $T_i$, and $U_{\phi i}$ profiles in both cases (Fig.~\ref{fig:1}).

The OMP profiles and velocities (Fig.~\ref{fig:1}) structures influence $E_r$ through the radial force balance constraint. Thus, to identify the origin of the difference between $E_r$ observed in Fig.~\ref{fig:2}, we consider the ion radial force balance, which holds in the full-$f$ GK formalism \cite{dif2009,frei2025}. Assuming isotropic ion pressure, the radial force balance reads \cite{hazeltine2013}:
$E_r  = \nabla_r P_i / (q_i n_i) + U_{i\phi} B_\theta - U_{i\theta} B_\phi$, with $E_r = - \nabla \phi_1 \cdot \hat{\bm e}_{\psi}$. As shown in Fig.~\ref{fig:1}, the diamagnetic ($\nabla_r P_i / [q_i n_i]$) and toroidal rotation ($U_{i\phi} B_\theta$) terms are nearly identical in both configurations and dominate $E_r$ for $\rho_{\mathrm{pol}} \lesssim 0.98$, implying a small contribution from the poloidal flow ($- U_{i\theta} B_\phi$). However, within the $E_r$ well ($\rho_{\mathrm{pol}} \sim 0.995$), they nearly cancel out, leaving $E_r \simeq -U_{i\theta} B_\phi$. Hence, the depth and the minimum value of the $E_r$ well ($E_{r,\mathrm{min}}$) are set by the amplitude and shape of the poloidal ion flow $U_{i\theta}$, which is much larger in amplitude in \textit{fav}. With the same magnetic field strength, the resulting inner $\bm{E}\times\bm{B}$ shear is approximately $1.6$ times stronger in \textit{fav} than in \textit{unfav} \cite{grover2024}.

The SOL $E_r$ agree well with experiments \cite{plank2023} and conduction-limited GK \cite{frei2025}, with only minor differences between \textit{fav} and \textit{unfav}. As a consequence, the outer $\bm{E}\times\bm{B}$ shear layer is $1.6$ times larger in \textit{fav} and exceeds the inner shear in amplitude in both \textit{fav} and \textit{unfav}. Altogether, Figs. \ref{fig:1} and \ref{fig:2} therefore indicate that the observed difference in the depth of the $E_r$ well (and its shear) cannot originate from NC effects, but instead arises from nonlinear interactions between turbulence and poloidal flow, as we demonstrate below.

\begin{figure}
\includegraphics[scale=0.45]{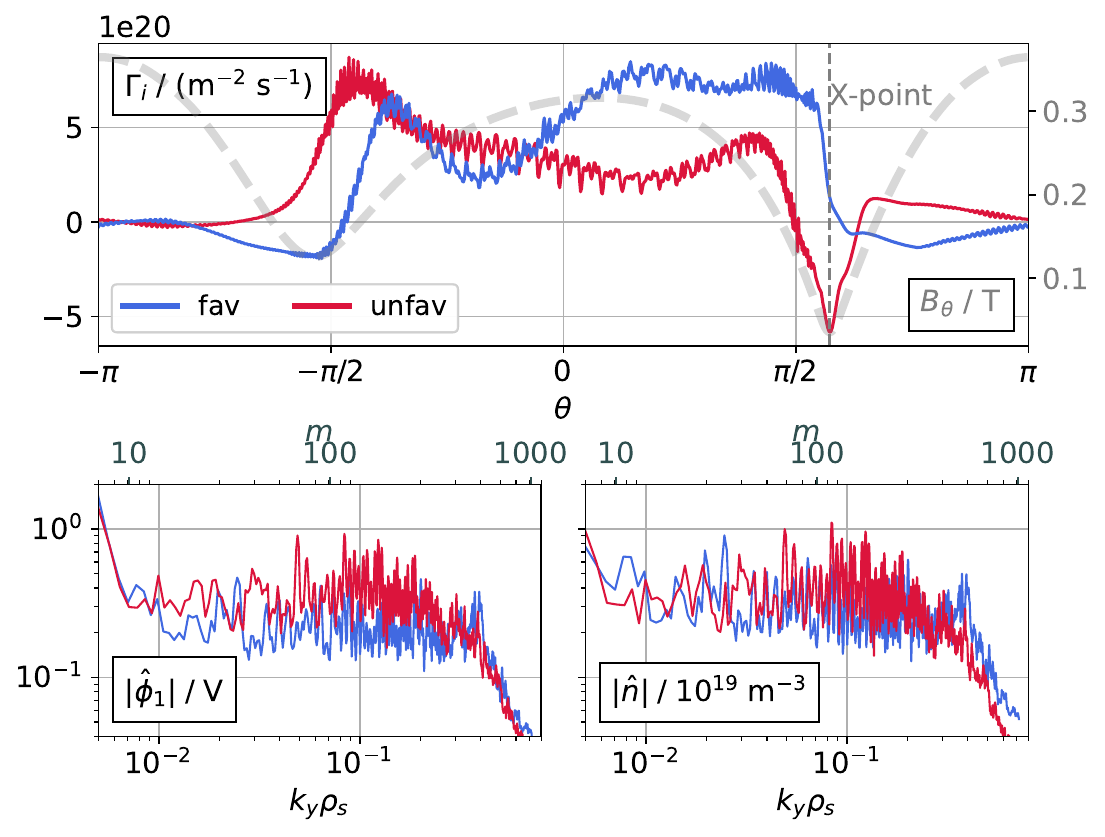} 
\caption{\label{fig:3} \textbf{Top:} Poloidal distribution of $\Gamma_i$ (time and toroidal averages) obtained from \texttt{GENE-X} vs. poloidal geometrical angle $\theta$ at $\rho_{\mathrm{pol}}=0.995$. \textbf{Bottom:} Fourier spectra of $\phi_1$ (left) and $n_e$ (right) vs. $k_y \rho_s$ ($\rho_s$ is the averaged ion sound Larmor radius) and poloidal mode number $m$.}
\end{figure}

\textit{Edge Transport and Turbulence.} We characterize here the edge transport and turbulence obtained from \texttt{GENE-X}. The top panel of Fig.~\ref{fig:3} compares the simulated poloidal distribution of the (toroidally and time-averaged turbulent) radial ion particle flux, $\Gamma_i = n_i u_\psi$ (with $u_\psi = \bm u \cdot \hat{\bm e}_\psi$ the radial component of the $\bm{E} \times \bm{B}$ velocity, $\bm u = c \bm E \times \bm B / B^2$) on the $\rho_{\mathrm{pol}} = 0.995$ flux-surface. In both configurations, $\Gamma_i$ exhibits fine structures predominantly around the OMP position on the LFS and decays on the high-field side. While $\Gamma_i$ features only moderate poloidal asymmetry, its magnitude is larger between the OMP and X-point position on the LFS in \textit{fav}. $\Gamma_i$ is also directed inwards (outwards) near the X-point on the LFS in \textit{unfav} (\textit{fav}), reflecting the contribution from $J^B$ (see \ref{sec:endmatter}). This poloidal profile of $\Gamma_i$ influences the poloidal distribution of the velocity stress $\Pi$ (proportional to $\Gamma_i$) and hence its FSA and gradient, as we show below.

The bottom plots of Fig.~\ref{fig:3} show the Fourier power spectra of $\phi_1$ and $n_e$ at $\rho_{\mathrm{pol}} = 0.995$. The spectra in \textit{unfav} exhibit higher amplitudes than in \textit{fav}. The temperature spectra show analogous trends \cite{bielajew2023}. Note that the phase-shifts characteristics remain qualitatively similar. Linear signatures of both eDW and TEM modes are found in the edge~\cite{bonanomi2024}, indicating comparable frequency–wavenumber relations in \textit{fav} and \textit{unfav}. Particle and heat transport crossing the separatrix are also of similar amplitudes. The present results demonstrate that the ion $\nabla B$ drift direction primarily influences the turbulence intensity being larger in \textit{unfav}. 

\begin{figure}
\includegraphics[scale=0.5]{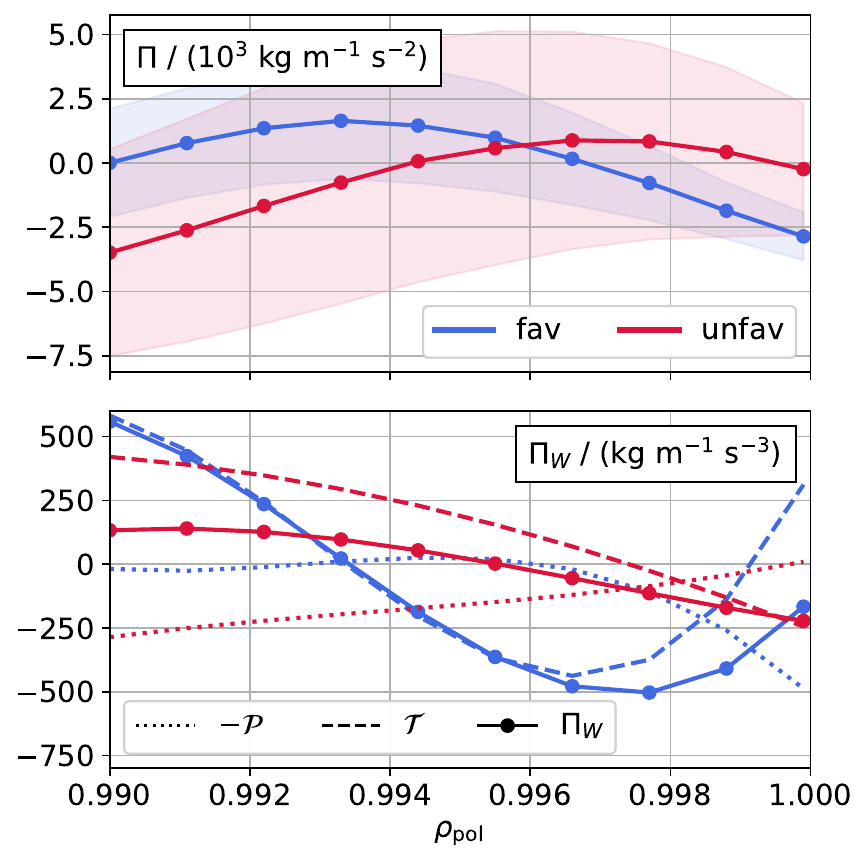} 
\caption{\label{fig:5} \textbf{Top:} Time-averaged radial profiles (with standard deviations) of $\Pi$. \textbf{Bottom:} Time-averaged radial profiles of the stress work $\Pi_W$ (solid), transport $\mathcal{T}$ (dashed), and negative production $-\mathcal{P}$ (dotted) contributions in \textit{fav} (blue) and \textit{unfav} (red).}
\end{figure}

\textit{Nonlinear Poloidal Flow Acceleration.} From Fig. \ref{fig:2}, we observe that $U_{i\theta}$ is dominated by the $\bm{E}\times \bm{B}$ velocity around the $E_r$ well, such that $U_{i\theta} \simeq u_\theta$ ($u_\theta= \bm u \cdot \hat{\bm e}_\theta$). Thus, an equation for the \textit{acceleration} of the FSA (zonal) poloidal flow momentum density, $M u_\theta$ (with $M=\sum_\alpha n_\alpha m_\alpha$ the total mass density), can be obtained from the GK vorticity equation, Eq. (\ref{eq:vorticity}), and is given in Eq. (\ref{eq:upolfavre}): 

\begin{align} \label{eq:upolfavre}
\frac{\partial}{\partial t} \left( \fs{M u_\theta}\right) =  - \frac{d  }{d r} \Pi  + \Lambda.
\end{align}
From Eq. (\ref{eq:upolfavre}), the poloidal flow acceleration is driven by the negative radial gradient of the total (electrostatic) velocity stress, $\Pi =  \fs{M} \favre{u_\theta u_\psi}$, with $\favre{f}=\fs{Mf}/\fs{M}$ the Favre average operator \cite{favre1965}. The roles of diamagnetic currents and GAMs are discussed in \ref{sec:endmatter}. We focus here on the drive due to $\Pi$. From Eq. (\ref{eq:upolfavre}), $\Pi$ drives poloidal flow acceleration (damping) if its gradient is negative (positive): Fig. \ref{fig:5} shows a clear larger negative radial gradient in \textit{fav} around the $E_r$ well. Note that $\Pi$ is related to $\Gamma_i$ through $\Pi \simeq \langle m_i \Gamma_i u_\theta \rangle$. In particular, it constitutes the full-$f$ generalization of the RS $\mathcal{R}$ \cite{held2018}: $\Pi$ can drive nonlinear poloidal flow acceleration, not only through $\partial_r \mathcal{R}$, but also through radial turbulent transport of poloidal momentum, poloidal density asymmetries, and triple fluctuation correlations. These effects, absent in previous formulations \cite{fedorczak2013,manz2018,grover2024}, remain finite for radially uniform $\mathcal{R}$. 

\begin{figure}
\includegraphics[scale=0.45]{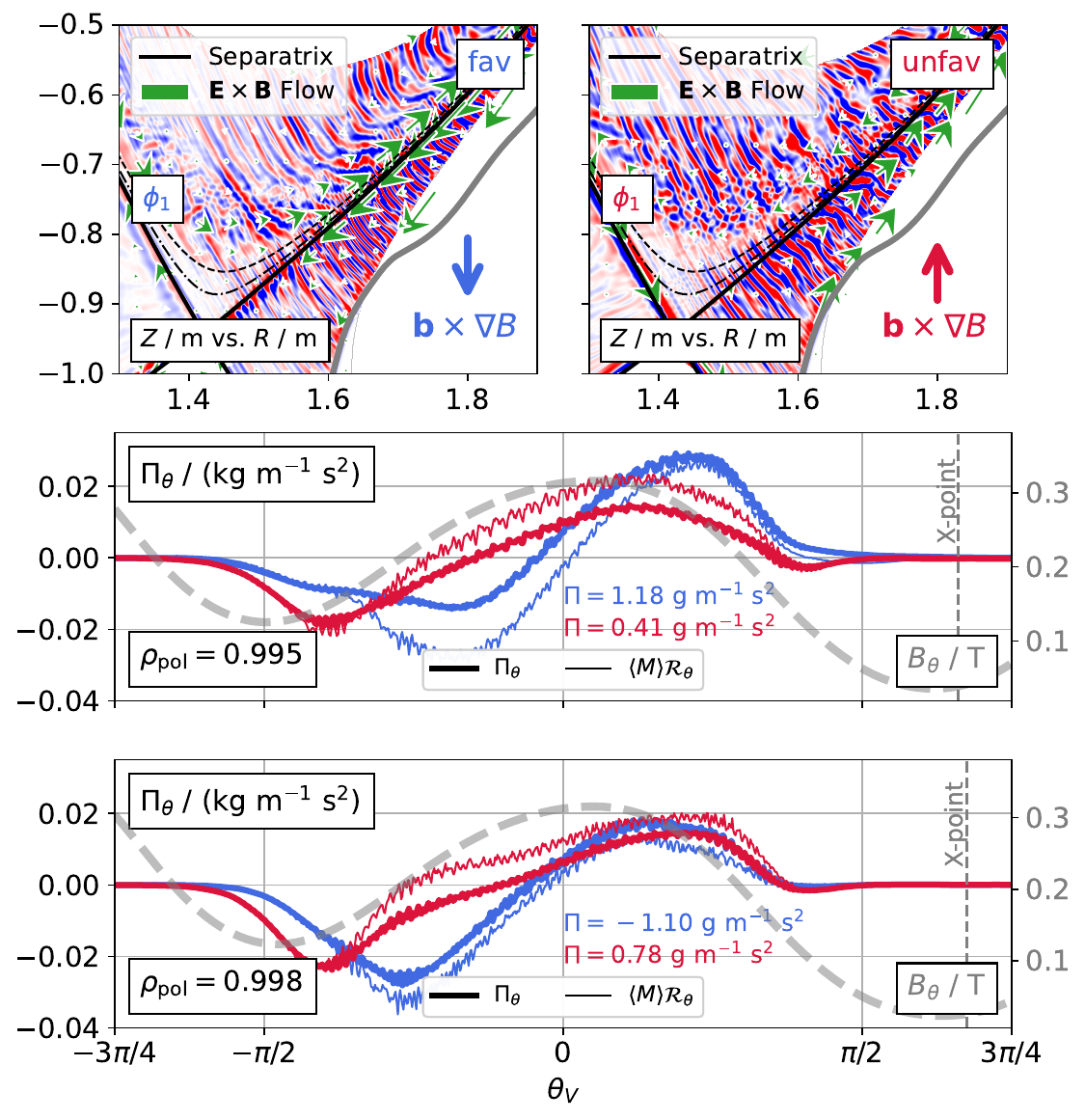} 
\caption{\label{fig:4} \textbf{Top:} Snapshots of $\phi_1$ and mean $\bm{E}\times\bm{B}$ flow (arrows) near the X-point in \textit{fav} (left) and \textit{unfav} (right). \textbf{Center and bottom:} Poloidal distributions of $\Pi_\theta$ (thick) and $\mathcal{R}_\theta$ (thin) at $\rho_{\mathrm{pol}}=0.995$ and $0.998$ vs.\ $\theta_V$  \cite{grover2024}. The X-point is indicated by the vertical dashed line.}
\end{figure}

For $\Pi$ to provide a finite contribution in Eq. \ref{eq:upolfavre}, radially dependent poloidal asymmetries in $\Pi$ must develop \cite{manz2018,grover2024}. These asymmetries can arise from the nonlinear interplay between magnetic-shear-induced (MSI) eddy tilting, X-point (or limiter) location, flow shear, and radial turbulent transport \cite{fedorczak2012,fedorczak2013,manz2018}. While the magnetic geometry fixes the MSI stress and X-point positions (identical in \textit{fav} and \textit{unfav}), the presence of a flow shear (inner or outer) layer can either nonlinearly amplify (\textit{fav}) or reduce (\textit{unfav}) the intrinsic MSI eddy tilting depending on the direction (sign of $B_\phi$) of the $\bm E \times \bm B$ flow \cite{fedorczak2012}. The top panels of Fig.~\ref{fig:4} show snapshots of $\phi_1$ fluctuations near the X-point. In \textit{unfav}, the $\bm{E} \times \bm{B}$ flow just inside the separatrix (in the outer shear) on the LFS is directed toward the X-point and remains weak. Thus, turbulent structures are radially stretched mainly due to the MSI \cite{manz2018}. In contrast, in \textit{fav}, the $\bm{E} \times \bm{B}$ flow is stronger near the separatrix and directed away from the X-point, shearing the turbulent structures and in the opposite direction (towards the OMP) in the outer shear layer, corresponding to a situation where the flow shear ($\alpha_u^{(u)}$) and MSI tilte ($\alpha_u^{(s)}$) angles are in the same direction, i.e. $\alpha_u^{(u)} \alpha_u^{(s)} >0 $ in \textit{fav} \cite{fedorczak2012,fedorczak2013,grover2024}. This localized $\bm{E} \times \bm{B}$ flow shear in the outer shear radially modulates the tilt of the turbulent eddies and the poloidal distribution of $\Pi$, thereby introducing a finite $\Pi$ gradient and yielding self-amplification. 

The middle and bottom panels of Fig. \ref{fig:4} show the time-averaged poloidal distribution of $\Pi_\theta$ ($\Pi = \int  d\theta_V  \Pi_\theta$) as a function of $\theta_V$ on two flux surfaces: inside the $E_r$ well and closer to the separatrix (outer shear layer). We also display the corresponding poloidal distribution $\mathcal{R}_\theta$ of the RS $\mathcal{R}$ to assess the importance of the full-$f$ approach. In \textit{unfav}, $\Pi_\theta$ exhibits weak radial variation and is poloidally asymmetric around the OMP ($\theta_V = 0$), such that its FSA remains constant on both flux-surfaces (small radial gradient). $\Pi_\theta$ closely follows $\mathcal{R}_\theta$, indicating that full-$f$ corrections are small in this case. Note that the poloidal asymmetries in $\mathcal{R}_\theta$ arise from the competition between flow shear and MSI eddy tilting \cite{grover2024}. In contrast, in \textit{fav}, $\Pi_\theta$ also displays a poloidal asymmetry around the OMP, but its amplitude varies radially: inside the $E_r$ well, $\Pi_\theta$ becomes negative with reduced amplitude above the OMP and positive with larger amplitude below, leading to a net positive value of $\Pi$. This asymmetry is strongly correlated with the poloidal distribution of $\Gamma_i$ (see Fig.~\ref{fig:3}). Near the separatrix, $\Pi_\theta$ remains asymmetric, but its amplitude becomes comparable on both sides of the OMP, yielding $\Pi < 0$. The radial variation of $\Pi$ induces a finite negative radial gradient of about $7$ larger in \textit{fav} than in \textit{unfav}. Noticeable deviations between $\Pi_\theta$ and $\mathcal{R}_\theta$ are observed in \textit{fav}, particularly inside the $E_r$ well, indicating non-negligible full-$f$ contributions in $\Pi$ (triple fluctuation correlation). The radial dependence of $\Pi_\theta$ in \textit{fav} modifies the radial profile (and gradient) of $\Pi$ shown in Fig.~\ref{fig:5}, thereby influencing the efficiency of the turbulence–mean flow interaction.

\textit{Nonlinear Energy Transfer.} To assess the efficiency of energy transfer from turbulence to the mean poloidal flow, we consider the FSA poloidal kinetic energy, $K_\theta$, whose evolution equation is given by Eq.~(\ref{eq:dkthetadt}): $K_\theta$ is driven by the negative of the work done by the stress $\Pi$, $\Pi_W$. $\Pi_W$ includes the contributions (see \ref{eq:stresswork}) from turbulent transport of mean kinetic energy ($\mathcal{T}$), and the energy production by turbulence $\mathcal{P}$ (see , such that $\Pi_W = \mathcal{T} - \mathcal{P}$ \cite{manz2012}. Fig. \ref{fig:5} shows the profiles of $\Pi_W$, and of $\mathcal{T}$ and $\mathcal{P}$ obtained from the simulations. $\Pi_W$ exhibits a substantially larger and negative amplitude around the $E_r$ well in \textit{fav} (see Fig. \ref{fig:2}). On the other hand, $\Pi_W$ decreases almost linearly toward the separatrix \textit{unfav} and remains much smaller than in \textit{fav}. These differences in $\Pi_W$ indicate a stronger nonlinear energy transfer from turbulence into $K_\theta$ in \textit{fav}. Fig. \ref{fig:5} shows $\mathcal{T}$ dominates within the $E_r$ well. In contrast, $-\mathcal{P}$ is dominant in the outer shear layer near the separatrix \cite{manz2018}. In \textit{unfav}, $\mathcal{T}$ and $\mathcal{P}$ have opposite signs and comparable amplitudes, but remain small in magnitude. Fig. \ref{fig:5} demonstrates that the energy transfer from turbulence to mean flow (mediated by $\Pi_W$) provides the poloidal flow acceleration near the separatrix in \textit{fav}. 

 \textit{Conclusions.} First-principles GK simulations of edge and SOL turbulence in realistic diverted geometry reproduce the experimentally observed difference in $E_r$ well depth between the \textit{fav} and \textit{unfav} configurations at AUG \cite{plank2023}. The simulations show that this difference arises from nonlinear turbulence-driven poloidal flows, which set both the depth and shear of the $E_r$ well. We demonstrate that the efficiency of this nonlinear energy transfer is enhanced (\textit{fav}) or reduced (\textit{unfav}) depending on the direction of localized $\mathbf{E}\times \mathbf{B}$ shear near the LFS separatrix. From the evolution equation of the mean poloidal kinetic energy, the strength of the turbulence-mean flow energy transfer is identified to be driven by the radial gradient of the full-$f$ velocity stress $\Pi$. Finally, turbulence activity is found to be larger in \textit{unfav}. These results provide the first self-consistent GK explanation of the drift configuration dependence of the $E_r$ well depth, identifying turbulence-mean flow energy transfer (mediated by the gradient of $\Pi$) as the key nonlinear mechanism yielding the deeper $E_r$ well in \textit{fav}. With the stronger turbulence in \textit{unfav}, this nonlinear mechanism driven by the $\bm E \times \bm B$ drift direction between the OMP and X-point on the LFS may underlie the observed configuration dependence of the H-mode (and I-mode) power threshold~\cite{diamond2005,cavedon2024}. Finally, flux-driven (fixed input power) simulations are required to further test this mechanism against experimental power thresholds. Finally, we note that similar trends are also observed in simulations of other devices, such as TCV \cite{rienacker2025}, indicating the universality of the present findings.

\textit{Acknowledgments.} The authors would like to thank B. De Lucca, S. Rienaecker, L. Vermare, P. Hennequin for useful discussions. This work has been carried out within the framework of the EUROfusion Consortium, funded by the European Union via the Euratom Research and Training Programme (Grant Agreement No 101052200 — EUROfusion). Views and opinions expressed are however those of the author(s) only and do not necessarily reflect those of the European Union or the European Commission. Neither the European Union nor the European Commission can be held responsible for them.

\appendix
\section{End Matter} \label{sec:endmatter}

\textit{Magnetic Field Orientation.} The magnetic field $\mathbf{B}$ can written as $\mathbf{B} =  B_\phi \hat{\bm e}_\phi + B_\theta \hat{\bm e}_\theta$, with $\hat{\bm e}_\phi$ ($B_\phi$) and $\hat{\bm e}_\theta$ ($B_\theta$) the toroidal and poloidal unit vectors (components of $\bm B$). We choose $\hat{\bm e}_\phi$ to be such that $B_\phi > 0$ in both configurations. We invert \textit{only} the direction of $\hat{\bm e}_\phi$ between \textit{fav} and \textit{unfav}, while the poloidal magnetic field has the same direction. $\hat{\bm e}_\theta$ points away from the active X-point on the LFS in both configurations. This implies that the direction of the $\bm E \times \bm B$ inside the $E_r$ well (see Fig. \ref{fig:4}) is positive (negative) in \textit{fav} (\textit{unfav}). The unit radial vector is defined as $\hat{\bm e}_\psi = \nabla \psi / \left| \nabla \psi\right|$ ($\psi$ is the poloidal flux function), such that $\hat{\bm e}_\theta =  \sigma_B \hat{\bm e}_\phi \times \hat{\bm e}_\psi$ (with $\sigma_B = 1$ in \textit{fav} and $-1$ in \textit{unfav}). Finally, we note that, to facilitate comparison between the two configurations, the signs of the velocities in \textit{unfav} are adjusted to match that of \textit{fav}.

\textit{Simulation Setup.} We summarize here the numerical setup employed in \texttt{GENE-X}. While a comprehensive description of the numerical methods implemented in \texttt{GENE-X} can be found in \cite{michels2021,frei2025vspec}, only the most relevant aspects are highlighted here.
At the boundaries of the numerical domain (extending from $\rho_\mathrm{pol}=0.9$ to $1.03$), the distribution functions are assumed to be Maxwellian \cite{frei2025vspec}, with density and temperatures fixed at values close to experiments. The electromagnetic fields are set to zero. To avoid artificial flow shear near the edge boundary, Neumann boundary conditions on flux variables are applied. For the numerical resolution, we use $6$ parallel (Hermite) and $4$ perpendicular (Laguerre) spectral coefficients in velocity space \cite{frei2025}, $32$ poloidal planes, and a constant grid spacing of $\sim 1.4$~mm (about $1.5$ ion thermal gyroradius $\rho_i$ near the separatrix) within the poloidal plane.

To mimic the effects of neutral ionization in the edge, we introduce a poloidally homogeneous, axisymmetric, radially localized Gaussian energy-conserving density source centered at $\rho_{\mathrm{pol}}=0.99$, with a constant amplitude of $1.5\times10^{22} \mathrm{s}^{-1}$ for both \textit{fav} and \textit{unfav}. In the present setup, particle and energy sources and sinks arise from a combination of Dirichlet boundary conditions at the edge and SOL boundaries—where the profiles are fixed to values close to experiments—and perpendicular hyperdiffusion applied within a narrow buffer region between ghost points and the numerical domain. Upon discretization, this approach is equivalent to a Krook-like operator that relaxes the distribution function toward a Maxwellian with fixed density and temperature near the boundary. Acting as a particle and energy reservoir, this numerical strategy injects or removes particles and energy to enforce the boundary conditions. As a consequence, the present simulations can be considered effectively flux-driven, but with adaptive sources, while the profiles between the edge and SOL boundaries are free to evolve.

\textit{GK Vorticity Equation.} We provide a streamline derivation the GK vorticity equation, which we use to obtain the FSA poloidal flow acceleration, given in Eq.~\ref{eq:upolfavre}. Using the QN equation, $ \omega =\sum_\alpha q_\alpha n_\alpha$, with the electrostatic vorticity variable $\omega = \nabla \cdot \bm \Omega$ ($\bm \Omega =   c^2 M  \bm E /  B^2 $, $\bm E = - \nabla \phi_1$) and applying the FSA operator, $\fs{f} =  (V')^{-1} \oint d S f / \left|\nabla \psi \right|  := \int d \theta_V \fs{f}_\phi   / (2 \pi)$ (with $V(\psi)$ the volume enclosed by the $\psi$ flux-surface with $V' = d V / d \psi$, $dS$ the surface element of the flux-surface $\psi$, $\theta_V$ the flux-surface average angle defined in \cite{grover2024}, and $\fs{\cdot}_\phi$ the toroidal average), we derive, by taking the zeroth-order moment of the GK Vlasov equation, 

\begin{align} \label{eq:vorticity}
   \frac{\partial }{\partial t} \fs{\omega}  +  \frac{d}{dr}\fs{ J_\perp } =0,
\end{align}
\\
with $d / dr $ the radial derivative along $\hat{\bm e}_\psi$. Here, $J_\perp = \bm J_\perp \cdot \hat{\bm e}_\psi$ is the radial component of the \textit{total} and perpendicular \textit{gyrocenter} current density, $\bm J_\perp = \sum_\alpha  \bm J_{\perp \alpha} = \sum_\alpha q_\alpha \int d W  \dot{\bm R}_\perp f_\alpha$. Here, $\bm J_{\perp \alpha}$, can be separated into different contributions: $\bm J_{\perp  \alpha} = \bm J^{es}_{  \alpha} + \bm J^{em}_{ \alpha} + \bm J^{B}_{  \alpha} + \bm J_{\alpha}^{pol}$, with $\bm J^{es}_{  \alpha} = q_\alpha n_\alpha \bm u$ (electrostatic), $ \bm J^{em}_{  \alpha} = J_{\parallel \alpha} \bm B_1 /B$ (flutter with $J_{\parallel \alpha} = q_\alpha n_\alpha U_{\parallel \alpha}$ the parallel current), $\bm J^{B}_{  \alpha} =  c P_{\parallel \alpha} \nabla \times \bm b  / B   + c  P_{\perp \alpha} \bm b \times \nabla B / B^2  $ (diamagnetic), $\bm J_{ \alpha}^{pol} = -  c m_\alpha n_\alpha \bm b \times \nabla u^2 / (2 B)$ (polarization). Hereafter, we neglect $J^{em}$, which is associated with the Maxwell stress, due to the low values of electron plasma beta in the present cases. Note that in Eq. (\ref{eq:vorticity}), the contribution from the density source vanishes since no vorticity is injected. 

\begin{figure}
\includegraphics[scale=0.55]{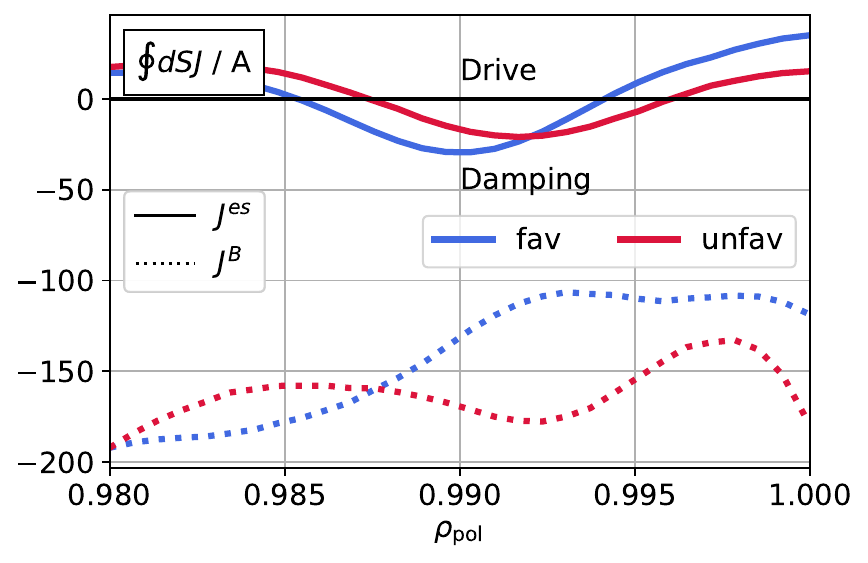} 
\caption{\label{fig:6} Surface-integrated and time-averaged radial profiles of total $J^{es} = \bm J^{es} \cdot \hat{\bm e}_\psi$ (solid) and $J^{B} = \bm J^{B} \cdot \hat{\bm e}_\psi$ (dotted). Positive (negative) values (relative to the direction of $u_\theta$) indicate drive (damping).}
\end{figure}

\textit{Radial Current Contributions.} We briefly discuss the contributions from the radial electrostatic ($J^{es} = \bm J^{es} \cdot \hat{\bm e}_\psi$) and diamagnetic ($J^{B} = \bm J^{B} \cdot \hat{\bm e}_\psi$) currents. These currents can either drive or damp the mean vorticity $\omega$ (mean poloidal flow) depending on their sign (relative to the poloidal flow velocity). $J^{es}$ typically acts as driving term through nonlinear interactions with turbulence, whereas $J^{B}$ generally damps poloidal flows. In particular, in the absence of turbulence, $J^{B}$ damps poloidal flows via parallel ion viscosity (NC damping \cite{kim1991,peeters2000}). In Fig.~\ref{fig:6}, the FSA and time-averaged radial profiles of $J^{B}$ and $J^{es}$ are shown in \textit{fav} and \textit{unfav}. Here, the signs are defined relative to $u_\theta$ such that positive (negative) values indicate drive (damping). The contribution from $J^{B}$ is negative in both cases, but its amplitude is larger in \textit{unfav} around the $E_r$ well, indicating stronger damping. In contrast, the electrostatic radial current $J^{es}$ is positive and stronger in the \textit{fav} case near the $E_r$ well, providing a stronger drive in \textit{fav}. The combination of $J^{B}$ and $J^{es}$ therefore suggests a stronger net drive of the poloidal flow in \textit{fav}. However, we note that the present analysis is qualitative since a rigorous balance analysis of Eq.~(\ref{eq:vorticity}) is hindered by the separation of characteristic time scales, with $J^{B}$ evolving on longer collisional time scales and $J^{es}$ on shorter turbulence time scales. Consequently, we focus here on the role of $J^{es}$ in driving mean poloidal flows.

\begin{figure}
\includegraphics[scale=0.55]{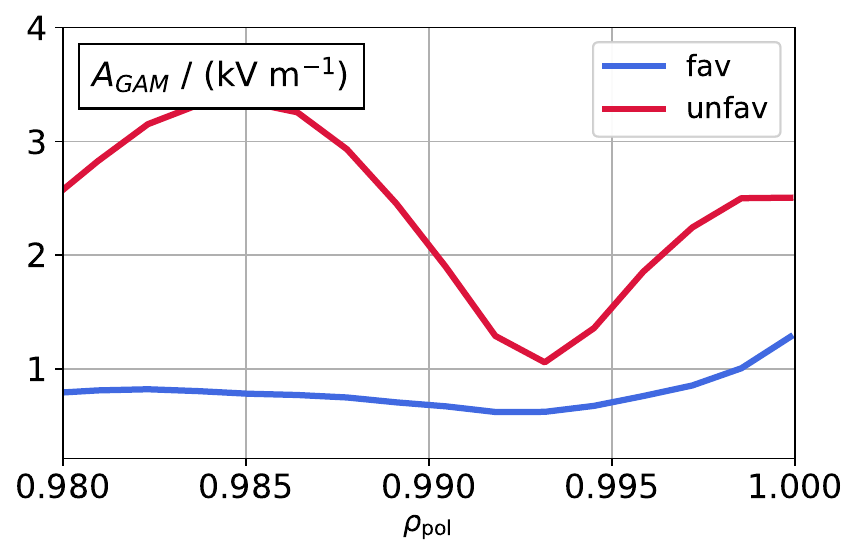} 
\caption{\label{fig:7} Radial profiles of the GAM amplitude ($A_{\textrm{GAM}}$) measured around the $E_r$ well in the frequency ($f$) band $10$~kHz~$\lesssim f \lesssim 30$~kHz.}
\end{figure}

\textit{GAM Oscillations.} The coupling between mean poloidal flows and diamagnetic currents can transfer energy to pressure sidebands via the excitation of GAMs \cite{scott2003,scott2005,conway2011}. These pressure sidebands can in turn couple to parallel currents and dissipate energy through electron resistivity or can be nonlinearly depleted \cite{scott2003}, thereby limiting the growth of mean (zonal) poloidal flows. In the present simulations, the GAM activity, which is inferred from the GAM amplitude $A_{\textrm{GAM}}$, is found to be stronger in \textit{unfav} around the $E_r$ well and reaches a minimum at the location of $E_{r,min}$, as shown in Fig.~\ref{fig:7}. In contrast, $A_{\textrm{GAM}}$ is much smaller in \textit{fav}. The difference in the GAM amplitude between \textit{fav} and \textit{unfav}, also evidenced in the standard deviations in Fig.~\ref{fig:5}, arises from nonlinear interactions with turbulence via the velocity stress $\Pi$. This is confirmed by a detailed Fourier analysis of $\Pi$ (not shown here) revealing a clear peak at the linear GAM frequency in \textit{unfav}. This enhanced GAM activity suggests a larger energy transfer through geodesic coupling in \textit{unfav}, which may further inhibit the buildup of mean poloidal flows and contribute to a shallower $E_r$ well in \textit{unfav}. A more quantitative analysis of GAM drive and interaction with turbulence \cite{conway2011} is deferred to future work.

\textit{Mean Poloidal Flow Acceleration Equation.} From the vorticity equation in Eq. (\ref{eq:vorticity}), an evolution equation for the mean (FSA) poloidal flow acceleration driven by turbulence can be derived. Using the QN condition in the electrostatic current, such that $J^{es} = u_\psi \nabla \cdot \bm \Omega$, and integrating over $\psi$, the \textit{acceleration} of the FSA poloidal flow momentum density ($M u_\theta$) is obtained and is given by Eq. (\ref{eq:upolfavre}). Note that the damping contribution of $J^B$ in Eq. (\ref{eq:upolfavre}) and the constant of integration are absorbed in $\Lambda$. The gradient of the velocity stress $\Pi$ appears due to the combination of the radial components of $J^{es}$ and $J^{pol}$ assuming that $\nabla \cdot \bm u \simeq 0$ and neglecting the variation of $B$. Finite Larmor radius corrections to $\Pi$ \cite{smolyakov2000,sarazin2021}) are absent here, but will be subject to future work. Note that the total stress $\Pi$ and the RS stress $\mathcal{R} = \fs{ \widetilde{u_\psi}\widetilde{u_\theta} }$ ($\widetilde{f} = f - \fs{f}$), are related by $\Pi   = \fs{M} \mathcal{R} + \fs{\widetilde{M} \widetilde{u_\theta} \widetilde{u_\psi}} + \fs{M} \left(\favre{u_\theta} \favre{u_\psi} -\favre{\widetilde{u_\theta}}\favre{\widetilde{u_\psi}} \right)$ \cite{held2018}.  

From Eq. (\ref{eq:upolfavre}), the evolution equation for the FSA poloidal kinetic energy, $K_\theta  =  \fs{ M u_\theta}^2 / 2$, is

\begin{equation}
\label{eq:dkthetadt}
\frac{1}{\fs{M}}\frac{\partial K_\theta}{\partial t} = -\Pi_W + \favre{u_\theta}\fs{\Lambda},
\end{equation}
Here, $\Pi_W = \favre{u_\theta} d_r \Pi$ denotes the stress work associated with $\Pi$ and can be written as \cite{manz2012}

\begin{align} \label{eq:stresswork}
\Pi_W & =  \frac{d}{dr} \left(\favre{u_\theta}\Pi\right) - \Pi \frac{d}{dr} \favre{u_\theta} \nonumber \\
& := \mathcal{T} - \mathcal{P},
\end{align}
with $\mathcal{T}$ representing the turbulent transport (flux) of mean kinetic energy and $\mathcal{P}$ the energy production by turbulence \cite{manz2012}. Note that the sign of $K_\theta$ and $\Pi_W$ is invariant between \textit{fav} and \textit{unfav}, being quadratic in $u_\theta$.

\nocite{*}

\bibliography{apssamp}

\end{document}